\documentclass{optica-article}

\journal{opticajournal} 

\articletype{Research Article} 

\usepackage{lineno}

\usepackage{mathtools}
\usepackage{multirow}
\usepackage[capitalise]{cleveref} 
\usepackage{xcolor}               
\usepackage{tabularray}
\UseTblrLibrary{booktabs}
\definecolor{tint}{HTML}{F3EFE7}  

\begin{document}

\title{Hybrid integration of a SESAM with a silicon nitride photonic integrated circuit}

\author{Amaryllis Verstraete,\authormark{1,\dag} Tom Reep,\authormark{1,\dag} Jan de Haan,\authormark{1,\dag} Eero Koivusalo,\authormark{2} Riku Isoaho,\authormark{2} Thi Ngoc Lam Tran,\authormark{1} Bart Kuyken,\authormark{1} and Kasper Van Gasse\authormark{1,*}}

\address{\authormark{1}Photonics Research Group, Ghent University - imec, B-9052 Gent, Belgium\\
\authormark{2}RefleKron, FI-33720 Tampere, Finland\\
\authormark{\dag}These authors contributed equally to this work.}

\email{\authormark{*}kasper.vangasse@ugent.be} 

\begin{abstract*}
The realization of high-performance femtosecond lasers on photonic integrated circuits (PICs) requires robust, low-loss coupling between a saturable absorber and a photonic waveguide. 
In this work, we demonstrate the hybrid integration of a semiconductor saturable absorber mirror (SESAM) with a silicon nitride (SiN) PIC by micro-assembly.
This micro-assembly strategy transfers the non-linear functionality of the absorber onto the PIC.
We characterize the linear and non-linear performance of an unbonded SESAM and of a SESAM bonded to inverse-tapered SiN waveguides.
At 1560~nm, the bonded device shows a waveguide-coupled modulation depth of 5.2\%, a maximum reflectivity of 37.2\%, and a saturation fluence of 47~\textmu J/cm$^2$. 
This work demonstrates that SESAMs can be integrated directly onto silicon nitride-based PIC platforms and provides a pathway towards fully integrated, self-starting, chip-scale ultrafast lasers.
\end{abstract*}

\section{Introduction}
Femtosecond laser systems and optical frequency combs are essential for a wide range of applications, including dual-comb ranging \cite{Camenzind22, chen2025absolute}, ultra-low noise microwave generation \cite{fortier2011generation, xie2017photonic}, dual-comb spectroscopy \cite{han2024dual, picque2026dual}, and optical atomic clocks \cite{newman2019architecture, wu2025vernier}. 
While these systems offer excellent performance, they rely on fiber or free-space optical components. This results in a relatively large footprint and cost, limiting their widespread deployment.
Replacing such table-top femtosecond laser systems with photonic integrated circuits (PIC) would offer significant advantages in terms of cost, power consumption, and mechanical stability.

However, chip-scale femtosecond laser systems and optical frequency comb generators have mainly been limited to lab-scale demonstrations, with different integration approaches presenting trade-offs in performance, complexity, and scalability \cite{chang2022integrated}.
For example, heterogeneously and micro-transfer printed semiconductor mode-locked lasers achieve self-starting mode-locking with picosecond pulse durations \cite{wang2017iii, cuyvers2021low} but have limited pulse energy.
Moreover, dissipative Kerr soliton microcombs can generate broadband combs with sub-picosecond pulse duration \cite{helgason2023surpassing, xiang2021laser, brasch2016photonic} but also provide limited pulse energies.
Other approaches, such as a rare-earth doped SiN waveguide-based Mamyshev oscillator \cite{qiu_high-pulse-energy_2026}, have enabled high pulse energy but require a dedicated start-up procedure.

One of the main challenges in achieving stable on-chip mode-locking is the integration of a saturable absorber (SA) with low non-saturable loss, high modulation depth, and self-starting capabilities. 
For example, on-chip saturable absorbers have been demonstrated in the form of non-linear SiN waveguide interferometers, which have been used to demonstrate a Q-switched mode-locked laser at 1.9~\textmu m, achieving 215~fs pulses at gigahertz repetition rates \cite{Shtyrkova}. 
While CMOS-compatible, self-starting continuous-wave (CW) mode-locking with these non-linear interferometers has not been demonstrated to date.  

In contrast, semiconductor saturable absorber mirrors (SESAMs) are a type of saturable absorbers that offer reliable self-starting \cite{keller1996semiconductor}.
These devices consist of an epitaxially grown Bragg mirror combined with a saturable absorbing layer with a short recovery time, enabling sub-picosecond pulse generation in free-space and fiber lasers \cite{keller2003recent,honninger1999q}. 
Moreover, direct bonding of a SESAM to a monolithic crystalline laser cavity was used to demonstrate a compact, alignment-free, ultra-low-noise mode-locked laser operating at 1550 nm \cite{Shoji2016MonolithicLaser}.
Furthermore, SESAMs have enabled mode-locking in erbium-doped silica-on-silicon planar lightwave circuits, with the SESAM held against the waveguide facet by a micro-positioner \cite{Byun2009WaveguideLaser}, and more recently dispersion-managed MLLs in erbium-doped SiN waveguides \cite{li2026fullyintegrateddispersionmanagedfemtosecond}.

In this work, we demonstrate a robust micro-assembly approach that combines the performance of SESAMs with the scalability of SiN photonics. 
An indium phosphide (InP)-based SESAM is bonded directly to the polished edge facet of a SiN PIC using a UV-curable epoxy [\cref{fig:design_and_fab}(a)]. 
To ensure low-loss waveguide-SESAM coupling, an inverse waveguide taper expands the waveguide mode at the facet and reduces diffraction loss in the bonding gap.
We characterize the SESAM before and after bonding, using two SESAMs from the same wafer, and obtain a waveguide-coupled, unsaturated reflectivity of up to 61\% at 1600 nm, and a modulation depth of 5.2\% and a saturation fluence of 47~\textmu J/cm$^2$ at 1560 nm. 
Although demonstrated at 1550~nm and 1560~nm, the integration technique is platform-agnostic and can be applied to any edge-coupled PIC platform and to other wavelengths, providing a general route to on-chip saturable absorbers.

\section{Design and fabrication}
\begin{figure}[htbp]
  \centering
  \includegraphics[width=\textwidth]{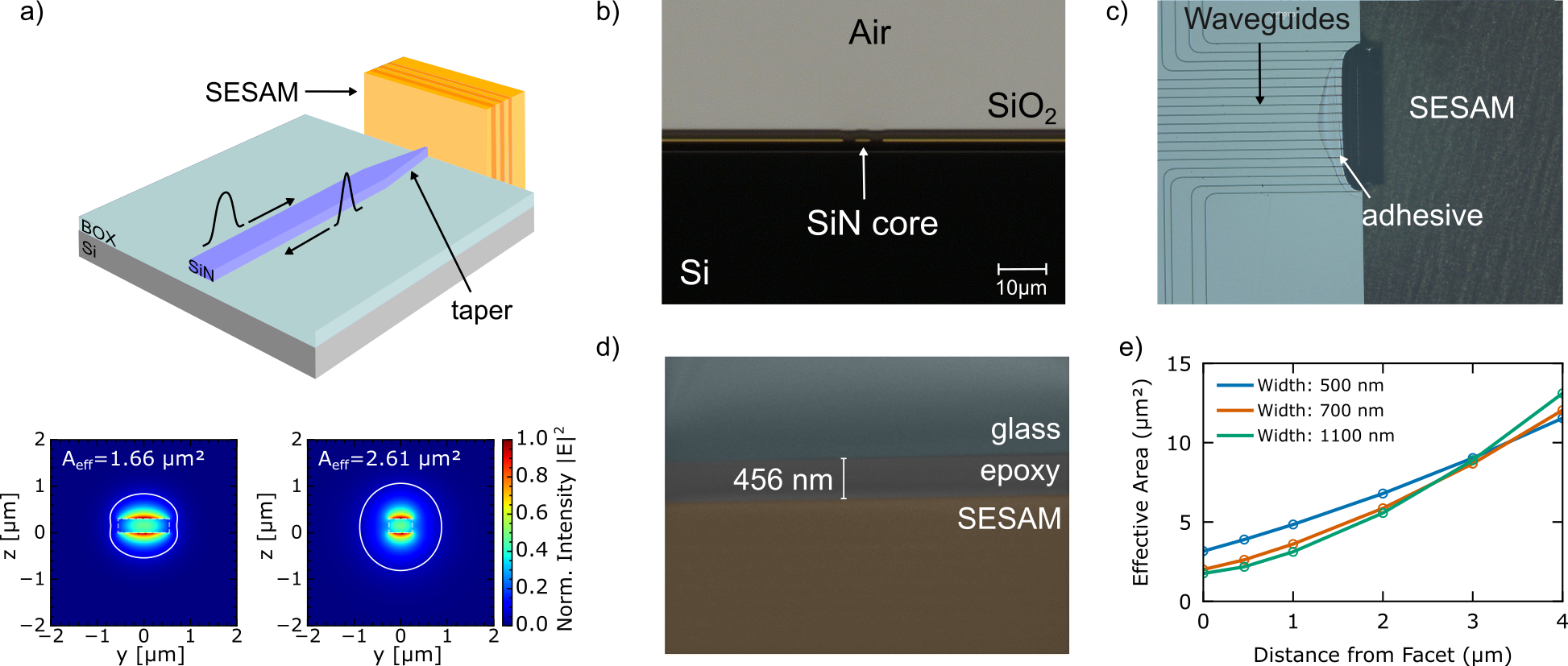}
  \caption{Design and integration of the SESAM-on-chip device. 
  (a) top: 3D schematic of the fabricated PIC with inverse taper and facet-bonded SESAM. The schematic also displays the temporal pulse shaping mechanism.
  bottom: simulated fundamental TM mode profiles of a 1.1~\textmu m wide waveguide (left) and 500~nm wide taper tip (right) at $\lambda=1550$~nm. White contours mark the effective mode area ($A_\text{eff}$). 
  (b) Dark-field optical microscope image of the polished facet, showing the SiN core embedded between a 3~\textmu m buried oxide (BOX) and 2~\textmu m top cladding. 
  (c) Top-view microscope image of a fabricated device.
  (d) False-colored SEM image of a FIB cross-section of a SESAM bonded to a glass substrate, showing an estimated bonding gap of 456~nm.
  (e) Simulated effective mode area ($A_\text{eff}$) as a function of distance from the facet for various taper tip widths.
  }
  \label{fig:design_and_fab}
\end{figure}

The primary challenge of this approach is low-loss coupling between the single-mode waveguide and the SESAM. 
Standard low-loss SiN waveguides provide high optical confinement, resulting in a small mode field diameter [\cref{fig:design_and_fab}(a)], which leads to rapid divergence of the light emitted from the facet. 
Even across a sub-micron bonding gap, this divergence reduces the overlap of the light reflected by the SESAM with the waveguide mode. 
We therefore designed inverse waveguide tapers that expand the mode at the facet and reduce the divergence-induced coupling loss. 
Tapers were designed with tip widths of 760~nm and 560~nm, resulting in fabricated tip widths of 700~nm and 500~nm as measured by scanning electron microscopy (SEM). 
Narrowing the tip reduces the confinement of the fundamental mode [\cref{fig:design_and_fab}(a)] and increases the effective mode area ($A_\text{eff}$) from 1.66~\textmu m$^2$ (width = 1100 nm) to 1.84~\textmu m$^2$ (width = 700 nm) and 2.61~\textmu m$^2$ (width = 500 nm) at $\lambda=1550$~nm. 
The tip width therefore provides a design degree of freedom to set the mode area and with it the fluence on the SESAM. 
A narrower tip couples more light back into the waveguide, but its larger mode area lowers the fluence and raises the pulse energy needed to saturate the absorber.
In this work, the simulations and experimental characterization were performed for the fundamental TM mode of the silicon nitride waveguide, consistent with, for example, high-power Tm:AlO$_x$ amplifiers realized on silicon nitride PICs \cite{Singh2025}. 
However, the approach is not limited to TM polarization and can be applied to TE operation without modification of the micro-assembly process.

The SiN waveguides were fabricated on a wafer consisting of a 300~nm LPCVD SiN layer on a silicon substrate with a 3~\textmu m thermally grown buried oxide (BOX).
The waveguides were defined by electron-beam lithography (EBL) and transferred into the SiN layer by a CF$_\text{4}$/H$_\text{2}$-based reactive ion etch (RIE). 
A 2 \textmu m thick silicon dioxide (SiO$_2$) top cladding was then deposited by inductively coupled plasma chemical vapor deposition (ICP-CVD).
Edge coupling facets were created by wafer-saw dicing, and were polished and further planarized by chemical-mechanical polishing (CMP), resulting in flat and low roughness facets suitable for precision bonding [\cref{fig:design_and_fab}(b)].

The SESAM consists of a distributed Bragg reflector (DBR), a quantum-well absorber layer and a dielectric protective layer.
The SESAM die (width: 500~\textmu m, height: 1000 \textmu m, thickness: 150 \textmu m) was realized by the precision lapping and cleaving of an epitaxially grown SESAM wafer.
The DBR has a designed peak reflectivity of 98.9\% at 1560~nm and the absorber is optimized for operation at 1560~nm, with a manufacturer-specified, free-space modulation depth of 15\%, a saturation fluence of 45~\textmu J/cm$^2$, and a recovery time of 1.4~ps.

The SESAM was attached to the facet of the SiN PIC using a NanoGlue micro-assembly system. 
A vacuum tip, used for the high-precision picking and placing of semiconductor dies and micro-components, is used for the pickup and placement of the SESAM.  
The vacuum tip is mounted on a six-axis arm with force sensors and sub-micron alignment precision. 
The SESAM was first dipped in a UV-curable epoxy ($n=1.56$), aligned to the polished facet, pressed into contact with the facet to minimize the bonding gap, and cured under UV illumination [\cref{fig:design_and_fab}(c)]. 
Because the bonding gap of the actual device cannot be measured without destroying it, we bonded a SESAM from the same batch to a glass substrate using identical assembly parameters and made a cross-section by focused ion beam (FIB) milling. SEM imaging of the cross-section gives an estimated gap width of 456~nm [\cref{fig:design_and_fab}(d)].
In the remainder of this work, we use 456~nm as the estimated bonding gap for simulations and calculations of the fluence.
Since the epoxy in the bonding gap does not guide the light, the mode continues to expand between the taper tip and the SESAM [\cref{fig:design_and_fab}(e)], so the mode area on the absorber is larger than at the facet.
For the 456~nm gap, the simulated effective mode area at the SESAM surface is 3.84~\textmu m$^2$ and 2.57~\textmu m$^2$ at 1550~nm and 3.89~\textmu m$^2$ and 2.62~\textmu m$^2$ at 1560~nm for the 500~nm and 700~nm tips, respectively. These values are used below to convert pulse energy to fluence at the absorber. 

\section{Linear characterization}
\begin{figure}[htbp]
  \centering
  \includegraphics[width=\textwidth]{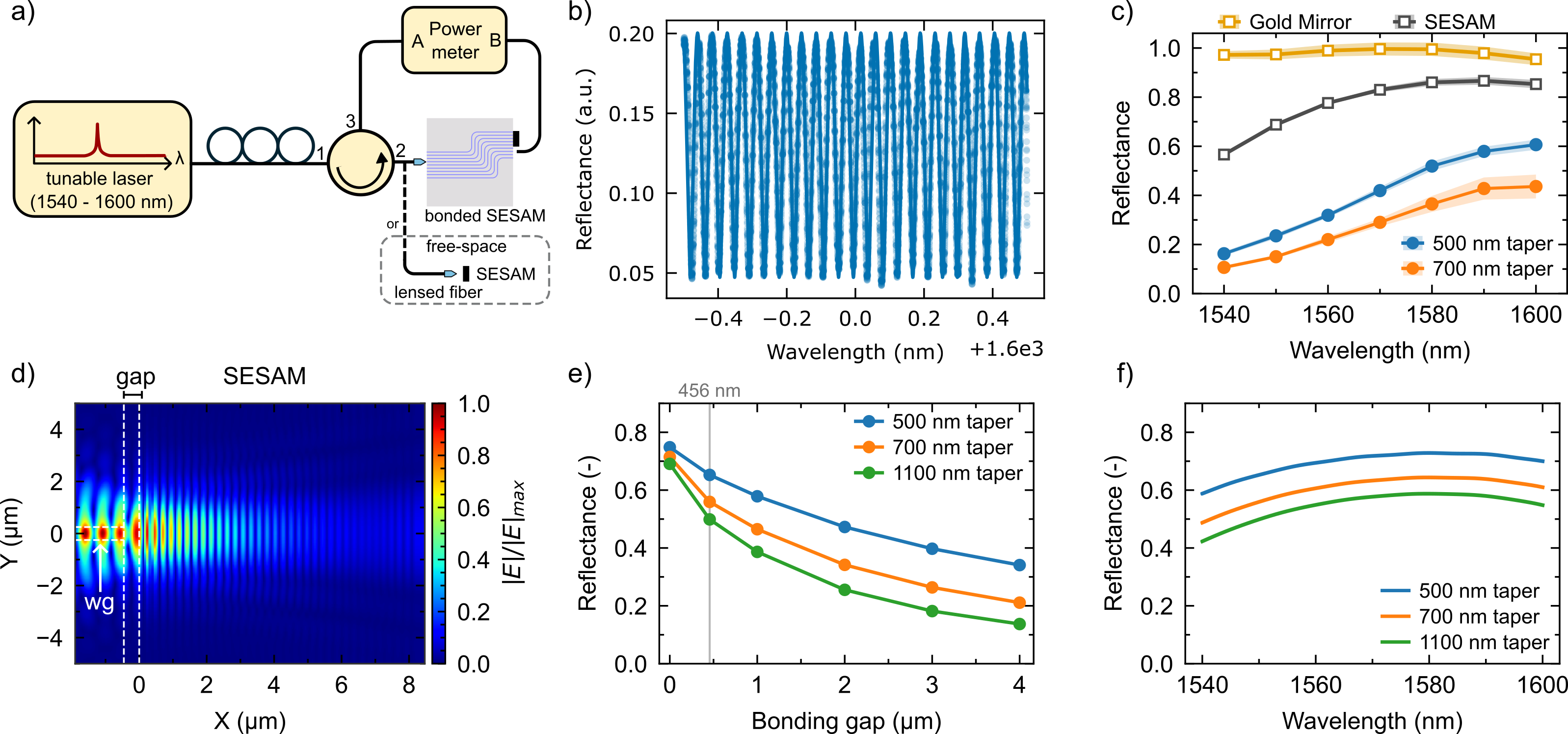}
  \caption{Linear characterization of a facet-bonded SESAM. 
  (a) Experimental setup: tunable CW laser, optical circulator and dual-channel power meter.
  (b) Fabry-P\'erot interference spectrum of the measured reflected power from a SESAM bonded to a 500~nm taper tip. 
  (c) Measured reflectivity of the gold mirror (yellow), bare SESAM (black), and bonded devices with 500~nm (blue) and 700~nm (orange) tips.
  (d) Simulated field for single-mode excitation of the facet-bonded SESAM with a 456~nm bonding gap at $\lambda=1550$~nm.
  (e) Simulated reflectivity into the fundamental mode of a lossless, fully saturated SESAM as a function of the bonding gap for the three tip widths at $\lambda=1550$~nm.
  (f) Simulated reflectivity into the fundamental mode of a fully saturated SESAM without non-saturable loss as a function of wavelength for a 456~nm gap and the three tip widths.}
  \label{fig:FP}
\end{figure}
We first determine the in-waveguide, unsaturated reflectivity of the bonded SESAM over a wavelength range of 1540 - 1600~nm with a low-power CW laser, using a de-embedded Fabry-P\'erot (FP) method \cite{walker1985SimpleAccurate}.
Light from a tunable CW laser is coupled into the SiN waveguide through an AR-coated lensed fiber, and an optical circulator directs the reflected signal to a power meter (\cref{fig:FP}(a)).
The input facet of the silicon nitride waveguide and the bonded SESAM form an on-chip FP cavity, producing periodic fringes in the reflection spectrum [\cref{fig:FP}(b)]. 
High-resolution sweeps ($\Delta\lambda=1$~pm) were recorded in narrow bands at 10~nm intervals.
To isolate the SESAM response, the fiber-to-chip coupling and facet reflectivity were first extracted from bare waveguides on the same chip. The propagation loss was determined from the intrinsic quality factors of ring resonators with varying coupling gap fabricated on the same platform. 
The bonded device data were then fitted in two steps. The phase and effective optical path length, which are robust against noise and power fluctuations, were extracted from the fringe periodicity. With these fixed, a least-squares fit determined the SESAM reflectivity.

To establish a baseline, the unmounted, bare SESAM from the same fabrication batch was measured in a free-space configuration, where the SESAM was placed in the focal spot of an AR-coated lensed fiber. 
This setup was calibrated with a gold mirror of known reflectivity (R=98\%, manufacturer specification). 
The measured gold mirror response was scaled to 98\% and the same scaling factor is applied to the SESAM data.

The results are summarized in \cref{fig:FP}(c).
Both the bare and bonded devices follow the wavelength dependence of the SESAM, with reflectivity increasing towards longer wavelengths. 
At 1540~nm the reflectivity of the 500~nm-tip device is 16\%, 3.5 times lower than the 57\% of the bare SESAM, but the difference reduces at longer wavelengths: at 1600~nm the 500~nm-tip device reaches 61\% versus 85\% for the bare SESAM, whose reflectivity peaks at 87\% around 1590~nm.
The 500~nm tip consistently outperforms the 700~nm tips, indicating that the lower confinement improves the overlap of the mode reflected from the SESAM.
Note that the bare measurement is a performance baseline rather than a direct reference for coupling loss: replacing air ($n \approx 1$) by the epoxy ($n = 1.56$) as the incident medium modifies the standing-wave field inside the SESAM stack. The bonded device reflectivity combines the modified material response with the coupling efficiency. 

To understand the dependence on tip width, we performed 3D finite-difference time-domain (FDTD) simulations using a commercial solver (Lumerical FDTD, Ansys). 
\Cref{fig:FP}(d) shows the top-view electric field at $\lambda=1550$~nm: the superposition of forward and reflected fields, as light couples from a 500~nm tip across the 456~nm epoxy gap into the SESAM stack, modeled as fully saturated and without non-saturable loss. 
The reflectivity coupled back into the fundamental mode as a function of gap and tip width [\cref{fig:FP}(e)] confirms that narrower tips give higher reflection and a weaker dependence on gap thickness, because the larger $A_\text{eff}$ at the facet reduces divergence in the adhesive layer and improves the overlap on return.
\Cref{fig:FP}(f) shows the simulated wavelength dependence for the measured 456~nm gap.

The measured reflectivities remain below the simulated values.
For the 500~nm tip at 1600~nm, the simulation predicts 77\% for a fully saturated absorber without non-saturable loss, while 61\% is measured for the unsaturated device. This is attributed to non-saturable loss in the SESAM, which is not included in the model, together with small angular misalignment and residual facet roughness.

\section{Non-linear characterization}

\begin{figure}[htbp]
  \centering
  \includegraphics[width=\textwidth]{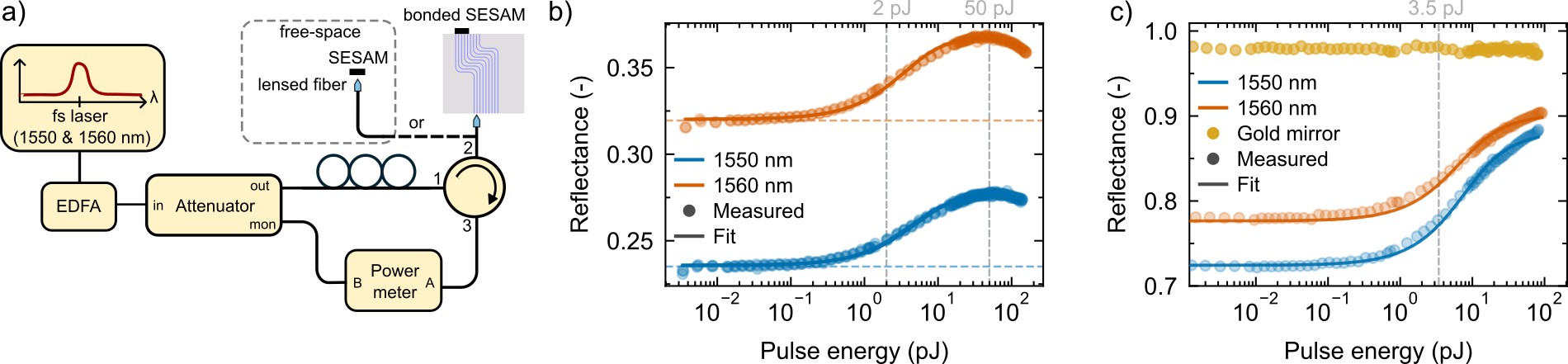}
  \caption{Saturation measurements of the bonded SESAM.
  (a) Setup for non-linear reflectivity measurements: pulsed fiber laser, erbium-doped fiber amplifier (EDFA), variable optical attenuator (VOA) with monitor port, and optical circulator.
  (b) Measured reflectivity versus incident pulse energy for the SESAM bonded to a 500~nm taper tip at 1550 and 1560~nm. Solid lines: fits of the saturable absorber model. Dashed lines: linear reflectivity of the bonded SESAM [\cref{fig:FP}(c)].
  (c) Same measurement for the bare SESAM in free space, together with the gold-mirror reference.
  }
  \label{fig:SA}
\end{figure}

To evaluate the saturable absorption required for mode-locking, we measured the fluence-dependent reflectivity of the bonded SESAM and extracted the modulation depth $\Delta R$, non-saturable reflectivity $R_\text{ns}$, and saturation fluence $F_{\text{sat}}$ \cite{haiml2004optical}. 
The setup [\cref{fig:SA}(a)] uses a pulsed fiber laser ($\tau = 5.6$~ps, $f_\text{rep} = 20$~MHz, $\Delta \lambda= 0.8$~nm) amplified by an erbium-doped fiber amplifier (EDFA).
A variable optical attenuator (VOA) with a monitor port tracks the input power, and the reflected signal is separated by a circulator and recorded on a dual-channel power meter. 
Measurements were performed at 1550 and 1560~nm, limited by the tuning range of the pulsed source.
For the bonded devices, the absolute reflectivity scale was set by matching the low-power, linear part of each curve to the reflectivity from the linear characterization [\cref{fig:FP}(c)].
The linear characterization provides higher accuracy and bypasses the need to separately quantify setup and propagation losses.
For the bare SESAM, the gold-mirror calibration described above was used. 
To convert incident pulse energy to fluence we use $A_\text{eff}$ at the SESAM surface, i.e. at the 456~nm plane: 3.84/2.57~\textmu m$^2$ (1550~nm) and 3.89/2.62~\textmu m$^2$ (1560~nm) for the 500/700~nm tips.
For the free-space benchmark, a Gaussian beam with $1/e^2$ radii of 1.35~\textmu m $\times$ 1.40~\textmu m ($A_{\text{eff}}=5.9$~\textmu m$^2$) was assumed, as determined by fiber-to-fiber beam profiling. 

The reflectivity versus incident pulse energy [\cref{fig:SA}(b)] shows the onset of saturation above $\approx 2$~pJ and, above 50~pJ a roll-over consistent with two-photon absorption (TPA) in the InP-based SESAM layers.
The data were fitted with the model of Ref.~\cite{haiml2004optical}, which includes fluence-dependent saturation and TPA induced roll-off: 
\begin{equation}
R(F) = R_\text{ns}\, \frac{\ln\left[1+\frac{R_\text{lin}}{R_\text{ns}}\left(e^{F/F_{\text{sat}}}-1\right)\right]}{F/F_{\text{sat}}}\, e^{-F/F_2},
\label{eq:sat}
\end{equation}
where $R_\text{lin}$ is the low-fluence reflectivity, $R_\text{ns}$ the non-saturable (fully saturated) reflectivity, $F_{\text{sat}}$ the saturation fluence, and $F_2$ the roll-off parameter, with $\Delta R = R_\text{ns}-R_\text{lin}$.

\Cref{tab:sa} summarizes the extracted parameters before and after bonding.
At 1550~nm, $R_\text{ns}$ decreases from 87.6\% in free-space to 28.0\% for the 500~nm tip and 17.6\% for the 700~nm tip, and $\Delta R$ from 15.5\% to 4.4\% and 2.6\%.
At 1560~nm, where the SESAM reflectivity is higher, the 500~nm tip reaches $R_\text{ns}=37.2\%$ and $\Delta R=5.2\%$, versus 25.4\% and 3.4\% for the 700~nm tip.
Importantly, the reduction of $\Delta R$ is a scaling by the coupling efficiency rather than a degradation of the absorber. The relative modulation depth $\Delta R / R_\text{lin}$ is 0.22 (1550~nm) and 0.16 (1560~nm) in free-space, and 0.19 and 0.16 after bonding to the 500~nm tip, and 0.17 and 0.15 for the 700~nm tips.
Equivalently, the extinction ratio between the unsaturated and saturated states, $10\log_{10}(R_\text{ns}/R_\text{lin})$ changes only from 0.85 to 0.76~dB (1550~nm) and from 0.66 to 0.67~dB (1560~nm).

The saturation energy $E_\text{sat}$ decreases from 3.8~pJ (lensed fiber) to 2.3~pJ (500~nm) and 2.5~pJ (700~nm) at 1550~nm, and from 3.5~pJ to 1.8 and 2.0~pJ at 1560~nm, as expected from the smaller mode area on the SESAM in the bonded configuration. 
The corresponding saturation fluences, 60 and 47~\textmu J/cm$^2$ for the 500~nm tip and 96 and 75~\textmu J/cm$^2$ for the 700~nm tips, are of the same order as the free-space values of 63.9 and 58.7~\textmu J/cm$^2$ and the manufacturer specification of 45~\textmu J/cm$^2$ ($\lambda=1560$~nm). 
An exact agreement is not expected. Replacing air by epoxy as the incident medium changes the standing-wave pattern inside the quantum-well layers and therefore the effective saturation fluence of the absorber, and the fluence conversion relies on a simulated mode area at a bonding gap measured on a separate test sample. 

\begin{table}[htbp]
\caption{Saturable-Absorber Parameters of the Bare (Free-Space) and Bonded SESAM$^a$}
\label{tab:sa}
\centering
\begin{tblr}{
  colspec = {l | l c c c c c c|},
  row{2,4,6} = {bg=tint},
  cell{2,4,6}{1} = {bg=white},
  rowsep = 2pt, colsep = 7pt,
  vline{1,Z} = {0.08em},
}
\toprule
Device & $\lambda$ (nm) & $R_\text{lin}$ (\%) & $R_\text{ns}$ (\%)
 & $\Delta R$ (\%) & $\Delta R/R_\text{lin}$ & $E_\text{sat}$ (pJ)
 & $F_\text{sat}$ (\textmu J/cm$^2$) \\
\midrule
\SetCell[r=2]{l} Free space & 1550 & 72.1 & 87.6 & 15.5 & 0.22 & 3.8 & 63.9 \\
 & 1560 & 76.7 & 89.2 & 12.5 & 0.16 & 3.5 & 58.7 \\
\SetCell[r=2]{l} 500 nm tip & 1550 & 23.5 & 28.0 & 4.4 & 0.19 & 2.3 & 60.2 \\
 & 1560 & 31.9 & 37.2 & 5.2 & 0.16 & 1.8 & 47.4 \\
\SetCell[r=2]{l} 700 nm tip & 1550 & 15.0 & 17.6 & 2.6 & 0.17 & 2.5 & 95.6 \\
 & 1560 & 22.0 & 25.4 & 3.4 & 0.15 & 2.0 & 75.3 \\
\bottomrule
\end{tblr}
\raggedright
$^a$Bonded $R_\text{lin}$ values are taken from the linear characterization.
\end{table}

\section{Discussion}

\begin{figure}[htbp]
  \centering
  \includegraphics[width=0.5\textwidth]{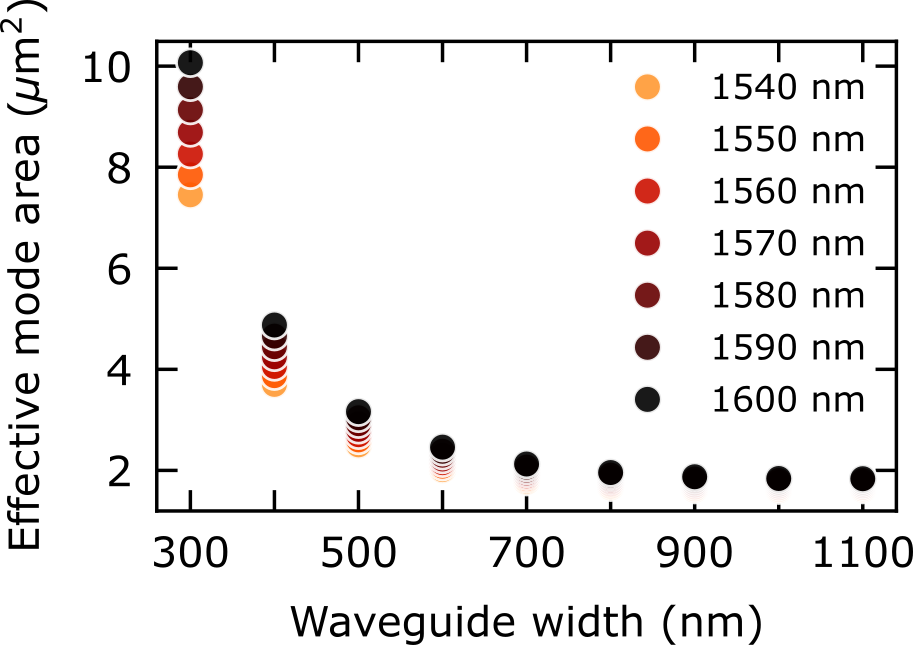}
  \caption{Simulated effective mode area ($A_\text{eff}$) of the fundamental TM mode as a function of waveguide width at the taper tip, for wavelengths from 1540 to 1600~nm.}
  \label{fig:A_effs}
\end{figure}

We have demonstrated the integration of a SESAM onto a SiN PIC by a micro-assembly process using a commercially available high-precision vacuum pick-and-place system. 
Moreover, we have experimentally shown that the micro-assembly process maintains the SESAM performance and that engineering of the silicon nitride waveguide facet allows to control parameters such as loss and saturation energy.
Bonded to a 500~nm inverse taper tip, the SESAM shows an in-waveguide modulation depth of 4.4\% (1550~nm) and 5.2\% (1560~nm), and a linear unsaturated reflectivity of up to 61\% (1600 nm), while the relative modulation depth of the free-space SESAM is preserved.
This performance is similar to that reported for a SiN non-linear interferometer (NLI) with a measured modulation depth of 9\% and a simulated non-saturable loss of approximately 4 dB \cite{Shtyrkova}, while providing self-starting capability and without the need for the 9 mm long non-linear waveguides.
Furthermore, simulations indicate that a thinner bonding gap would raise the coupled reflectivity [\cref{fig:FP}(e)] and that a narrower tip can further expand the mode to increase the saturation pulse energy.
An overview of simulated effective mode area as a function of waveguide width is shown in \cref{fig:A_effs}. 
These simulations show that the effective mode area can be increased from 2.7 to 8.3~\textmu m$^2$ by reducing the taper width, corresponding to an increase in saturation pulse energy from 1.8 to 5.6~pJ. 

Furthermore, the measured non-linear performance of the waveguide-coupled SESAM demonstrated in this work (\cref{fig:SA}) is directly compatible with the 30~pJ intracavity pulse energy reported for the silica waveguide-based mode-locked laser in \cite{Byun2009WaveguideLaser}.
This indicates that the performance of the reported device can directly support on-chip mode-locked laser operation.
More recently, the demonstration of a SESAM-based mode-locked erbium-doped SiN waveguide laser further confirmed the potential of SESAMs for self-starting femtosecond pulse generation in integrated laser cavities \cite{li2026fullyintegrateddispersionmanagedfemtosecond}. There, the SESAM was butt-coupled and co-packaged with the PIC, but the physical SESAM-waveguide interface and its waveguide-coupled linear and non-linear performance were not characterized, nor was a reproducible process for permanent attachment and alignment described.

Moreover, the micro-assembly approach is compatible with any photonic platform featuring edge couplers and can be directly combined with recently demonstrated rare-earth doped waveguide amplifiers, including erbium-doped SiN and Al$_2$O$_3$ \cite{Liu2022ErSiNAmplifier, GarciaBlanco2025ErAlOx}, ytterbium-doped Al$_2$O$_3$ for high-power ultrafast applications \cite{Zeng2026YbAlOx}, neodymium-doped Al$_2$O$_3$ at shorter wavelengths \cite{Jongebloed2026NdAlOx}, and emerging thin-film titanium:sapphire-on-insulator amplifiers for broadband ultrafast gain \cite{Yang2024TiSaOI}.
Together with the parameters demonstrated here, this provides a route towards fully integrated, self-starting, chip-scale femtosecond lasers.

\begin{backmatter}
\bmsection{Funding}
Fonds Wetenschappelijk Onderzoek (1SAA726N); European Research Council (101078281).

\bmsection{Acknowledgment}
A.V. acknowledges support from the Research Foundation -- Flanders (FWO) through a doctoral fellowship under grant No.~1SAA726N. This project has received funding from the European Research Council (ERC) under the European Union's Horizon Europe research and innovation program through the Starting Grant LASIQ, grant agreement No.~101078281.

\bmsection{Disclosures}
The authors declare no conflicts of interest.

\bmsection{Data availability}
Data underlying the results presented in this paper are not publicly available at this time but may be obtained from the authors upon reasonable request.

\end{backmatter}

\bibliography{refs} 

\end{document}